\documentstyle[12pt]{article}
\advance\voffset by -0.8in
\advance\hoffset by -0.55in
\input mssymb
\begin{document}
\newcommand{\V}{{\cal V}}
\newcommand{\Epsilon}{{\cal E}}
\newcommand{\Monster}{{\Bbb M}}
\newcommand{\reg}[1]{(\ref{#1})}
\newcommand{\sqr}[2]{{{\vcenter{\vbox{\hrule height.#2pt
\hbox{\vrule width.#2pt height#1pt \kern#1pt
\vrule width.#2pt}
\hrule height.#2pt}}}}}
\renewcommand{\square}{{\mathchoice\sqr{17}4\sqr{17}4\sqr{13}3\sqr{13}3}}
\newcommand{\Hil}{{\cal H}}
\newcommand{\C}{{\cal C}}
\newcommand{\ze}{{\Bbb Z}}
\newcommand{\re}{{\Bbb R}}
\newcommand{\ce}{{\Bbb C}}
\newcommand{\Aut}{{\rm Aut}}
\renewcommand{\C}{{\cal C}}
\title{Ternary Codes and $Z\!\!\!Z_3$-Orbifold Constructions of
Conformal Field Theories\footnote{Talk presented at the ``Monster Bash",
Ohio State University, May 1993}}
\author{P.S. Montague\\
Department of Applied Mathematics and Theoretical Physics\\
University of Cambridge\\
Silver Street\\
Cambridge CB3 9EW\\
U.K.}
\maketitle
\begin{abstract}
We describe a pair of constructions of Eisenstein lattices from ternary codes,
and a corresponding pair of constructions of conformal field theories from
lattices
which turn out to have a string theoretic interpretation.
These are found to interconnect in a similar way to results for binary codes,
which
led to a generalisation of the triality structure relevant in the construction
of the
Monster module. We therefore
make some comments regarding a series of constructions
of $V^\natural$. In addition, we present a complete construction of the
Niemeier
lattices from ternary codes, which in view of the above analogies should prove
to be of great importance in the problem of the classification of self-dual
$c=24$
conformal field theories. Other progress towards this problem is summarised,
and
some comments arise from this discussion regarding the uniqueness of the
Monster conformal field theory.
\end{abstract}
\section{Introduction}
We will begin in section \ref{ze2} by defining what shall be meant by a
conformal field theory (cft). Essentially, the definition is that of the
``vertex algebras'' of Borcherds \cite{Borchvalg}, who was inspired in part
to write it down as an axiomatisation of the structure of the Monster
module $V^\natural$ introduced by Frenkel, Lepowsky and Meurman (FLM)
\cite{FLMbook,FLMacadsci,FLMproc}.

We shall then proceed to briefly describe
the construction of the self-dual cft's $\Hil(\Lambda)$ from an even
self-dual lattice $\Lambda$.
A similar construction of an even self-dual lattice $\Lambda_\C$ from
a doubly-even self-dual linear binary code $\C$ will be described, which
gives rise to various analogies between the three structures. These
analogies then inspire us to write down a second construction of a
self-dual cft $\widetilde\Hil(\Lambda)$ from $\Lambda$ from a second
construction $\widetilde\Lambda_\C$ of an even self-dual lattice from
$\C$. This turns out to be what physicists would call a $\ze_2$-orbifold.

We shall explicitly describe the vertex operators involved, which is one
of the main strengths of our work.

Initially, the connection with codes was regarded as just a useful
analogy in that, as described above, one can use this ``dictionary'' of
properties to translate statements in simple systems into conjectures
in cft.
However, we noticed that $\widetilde\Hil(\Lambda_\C)\equiv
\Hil(\widetilde\Lambda_\C)$, which shows a deep structure
played by the code, and this leads to some understanding of FLM's triality
involution on $V^\natural$, which will be briefly discussed.

Section \ref{terncodes} should be regarded in some sense as an aside
from the main stream of this work, as no direct application to cft's
and/or the Monster $\Monster$ has yet been found. In this section, we
construct all of the Niemeier lattices by a pair of constructions from
a self-dual ternary code of length 24, whereas the binary constructions
provide only 12 of the 24 Niemeier lattices.
Continuing our previous analogies, this would suggest that all cft's of
central charge 24 may be constructed from lattices by orbifolding. This
has yet to be demonstrated. Though not directly relevant to $V^\natural$,
there are some points of interest to the Monstrous community!

Inspired by the above, in section \ref{ze3} $\ze_3$-orbifolds of the
theories $\Hil(\Lambda)$ for $\Lambda$ Niemeier are investigated.
The natural structure from which to begin, analogous to a ternary code,
is a (complex) lattice over the ring of Eisenstein integers $\Epsilon$,
which admits a natural third order no-fixed-point automorphism (NFPA).
This is inherited by the cft $\Hil(\Lambda_R)$, where $\Lambda_R$ is the
equivalent (real) $\ze$-lattice, and the orbifold we take is with respect
to this automorphism. We find a similar picture to that in the binary case,
though the correct cft analogy to the results of section \ref{terncodes}
remains to be found.

We construct the vertex operators explicitly, using techniques valid
for twists of orders 5, 7 and 13 also. This, in particular, gives us
more constructions of $V^\natural$. This may help to further illuminate
its unique structure.

Finally, in section \ref{class}, we summarise recent results on the
classification of self-dual cft's of central charge 24. This will be
related to $V^\natural$ in the sense that all such cft's should be contained
in $V^\natural\otimes V_{1,1}$ (where $V_{1,1}$ is the two-dimensional
Lorenztian theory), in an analogous way to the construction of the Niemeier
lattices from $II_{25,1}$. Some comments regarding the uniqueness of the
Monster module arise as a consequence.
\section{$Z\!\!\!Z_2$-Orbifolds and Binary Codes}
\label{ze2}
\subsection{Conformal field theories}
We define a {\em cft} to consist of a Hilbert space $\Hil$ and a set $\V=
\{V(\psi,z):\psi\in\Hil\,, z\in\ce\}$ of vertex operators,
$V(\psi,z):\Hil\rightarrow \Hil$ (we shall ignore any pretense at using
formal variables) and a pair of states $|0\rangle$, $\psi_L\in\Hil$ such
that
\begin{itemize}
\item $V(\psi,z)V(\phi,w)=V(\phi,w)V(\psi,z)$, the {\em
locality axiom}. Note that the left hand side is strictly defined only for
$|z|>|w|$, and similarly for the right hand side, and so we are to interpret
this equality in the sense that on taking matrix elements of either side the
resulting meromorphic
functions are analytic continuations of one another. This axiom is
physically reasonable in that we are to interpret the vertex operator
$V(\psi,z)$ as representing the insertion of the state $\psi$ on to the
world sheet of a string at the point $z$. For a bosonic string theory, the
order of such operator insertions must clearly be irrelevant.
\item $V(\psi_L,z)=\sum_nL_nz^{-n-2}$, with
\begin{equation}
[L_m,L_n]=(m-n)L_{m+n}+{c\over{12}}m(m^2-1)\delta_{m,-n}\,,
\end{equation}
for some scalar $c$, known as the {\em central charge}.
\item $V(\psi,z)|0\rangle=e^{zL_{-1}}\psi$, the {\em
creation axiom}.
\end{itemize}
The remaining axioms are technicalities,
listed here only for the sake of completeness.
\begin{itemize}
\item $x^{L_0}$ acts locally with respect to $\V$, {\em i.e.}
$x^{L_0}V(\psi,z) x^{-L_0}$ is local with respect to all the vertex
operators in $\V$. \item The spectrum of $L_0$ is bounded below.
\item $|0\rangle$ is the only state annihilated by $L_0$, $L_{\pm 1}$, {\em
i.e.}
the only su(1,1) invariant state in the theory.
\item $V\left(e^{z^\ast L_1}{z^\ast}^{-2L_0}\psi,1/z^\ast\right)^\dagger$ is
local
with respect to $\V$, {\em i.e.} the theory has a {\em hermitian structure}
\begin{equation}
\label{hermitian}
V\left(e^{z^\ast L_1}{z^\ast}^{-2L_0}\psi,1/z^\ast\right)^\dagger=
V(\overline\psi,z)
\end{equation}
for some antilinear map $\Hil\rightarrow\Hil$.
\end{itemize}
Physicists would call this a chiral bosonic meromorphic
hermitian conformal field theory. Essentially, it is what Borcherds
would refer to as a vertex algebra.
\subsection{Construction of $\Hil(\Lambda)$}
This brief description of a construction of the cft $\Hil(\Lambda)$
from an even lattice $\Lambda$,
and also later its $\ze_2$-orbifold $\widetilde\Hil(\Lambda)$,
will give the essential flavour of the third
order construction to be discussed later.

In this case\cite{DGMtwisted},
the Hilbert space $\Hil$ is taken to be the Fock space built up
by the action of creation operators on momentum states $|\lambda\rangle$,
$\lambda\in\Lambda$ an even lattice of dimension $d$.
These satisfy
\begin{eqnarray}
[a^i_m,a^j_n]=m\delta_{m,-n}\delta^{ij} & 1\leq i,j\leq d &
m,n\in\ze\nonumber\\
a_m^i|\lambda\rangle=0\quad\forall\quad m>0 & 1\leq i\leq d &
\lambda\in\Lambda\nonumber\\
(p^i\equiv a^i_0)|\lambda\rangle=\lambda^i|\lambda\rangle & 1\leq i\leq d &
\lambda\in\Lambda\\
{a^i_n}^\dagger=a^i_{-n} & & \nonumber\\
e^{i\lambda\cdot q}|\mu\rangle=|\lambda+\mu\rangle & & \nonumber\\
\end{eqnarray}
and the vertex operator corresponding to the state
\begin{equation}
\psi=\prod^M_{a=1}a^{i_a}_{-n_a}|\lambda\rangle
\end{equation}
is
\begin{equation}
V(\psi,z)=:\left(\prod^M_{a=1}{i\over(n_a-1)!}{d^{n_a}\over dz^{n_a}}
X^{i_a}(z)\right)e^{i\lambda\cdot X(z)}:\sigma_\lambda\,,
\end{equation}
where the normal ordering $:\cdots:$ indicates that creation operators are
written to the left of annihilation operators,
\begin{equation}
X^i(z)=q^i-ip^i\ln z+i\sum_{n\neq 0}{a^i_n\over n}z^{-n}
\end{equation}
is the string field ({\em i.e.} the coordinates of the string -- this theory
turns out to
correspond to a bosonic string moving on the torus $\re^d/\Lambda$) and
\begin{equation}
\widehat\sigma_\lambda\equiv e^{i\lambda\cdot q}\sigma_\lambda\,,
\end{equation}
satisfying
\begin{equation}
\widehat\sigma_\lambda\widehat\sigma_\mu=(-1)^{\lambda\cdot\mu}
\widehat\sigma_\mu\widehat\sigma_\lambda\,,
\end{equation}
are cocycle operators which ensure that locality is satisfied.

For a cft $\Hil$, let
\begin{equation}
\chi_\Hil(\tau)={\rm tr}_\Hil q^{L_0-c/24}\,;\qquad q=e^{2\pi i\tau}\,.
\end{equation}
This is the {\em partition function}, and is invariant under $T:\tau\rightarrow
\tau+1$ up to a phase of $e^{\pi ic\over 12}$, since we can show from our
axioms
that the spectrum of $L_0$ is integral.

We find that for $\Lambda$ self-dual $\chi_{\Hil(\Lambda)}(\tau)$ is also
invariant under
$S:\tau\rightarrow-1/\tau$, and so (for $c\in 24\ze$) under $\Gamma={\rm PSL}
(2,\ze)$, the modular group. [This is actually a physical requirement for the
theory
to be well-defined on the torus parameterised by $\tau$ in the usual sense.]

So, we define a cft $\Hil$ to be {\em self-dual} if $\chi_\Hil(\tau)$ is
$S$-invariant.

This construction $\Lambda\mapsto\Hil(\Lambda)$ provides us with a
``dictionary", as mentioned in the introduction, between properties of lattices
and
cft's\cite{PGmer}.
\subsection{Construction of $\Lambda_\C$}
We also have a similar dictionary between binary codes and lattices, provided
by the construction of a lattice $\Lambda_\C$ from a binary code $\C$ by
\begin{equation}
\Lambda_\C={\C\over\sqrt 2}+\sqrt 2\ze^d\,,
\end{equation}
{\em e.g.} $\Lambda_\C$ is even for $\C$ doubly-even.
\subsection{Construction of $\widetilde\Lambda_\C$ and
$\widetilde\Hil(\Lambda)$}
A second, or ``twisted", construction
\begin{equation}
\widetilde\Lambda_\C=\Lambda_0(\C)\cup\Lambda_3(\C)\,,
\end{equation}
where
\begin{eqnarray}
\Lambda_0(\C)&=&{\C\over\sqrt 2}+\sqrt 2\ze^d_+\nonumber\\
\Lambda_1(\C)&=&{\C\over\sqrt 2}+\sqrt 2\ze^d_-\nonumber\\
\Lambda_2(\C)&=&{\C\over\sqrt 2}+{1\over 2\sqrt 2}\underline 1+
\sqrt 2\ze^d_{(-)^{n+1}}\\
\Lambda_3(\C)&=&{\C\over\sqrt 2}+{1\over 2\sqrt 2}\underline 1+
\sqrt 2\ze^d_{(-)^n}\nonumber\,,
\end{eqnarray}
of a lattice from a binary code for $d=8n$ ($\ze^d_+=\{x\in\ze^d:x^2={0\bmod
2}\}$,
$\ze^d_-=\{x\in\ze^d:x^2={1\bmod 2}\}$, $\underline 1=(1,1,\ldots,1)$) inspires
a second (twisted) construction of a cft from a lattice (in $8n$ dimensions).
This turns out
to correspond to a string moving on the orbifold
$\left(\re^d/\Lambda\right)/\ze_2$,
and so we refer to it as an orbifold construction\cite{DGMtwisted}.

The basic idea is that we have a $\ze_2$ symmetry of the theory
$\Hil(\Lambda)$,
induced by the reflection symmetry of $\Lambda$, and we project out by this and
add in a module for the resulting theory to restore self-duality. It turns out
in fact
that there are only two inequivalent such irreducible modules (a result still
to be
proven in general, but verified explicitly in a particular case by Dong
\cite{dongmoonrep,dongtwisrep}). One
clearly gives us back $\Hil(\Lambda)$, while the other gives us what we call
$\widetilde\Hil(\Lambda)$, the orbifold theory.

Explicitly, the vertex operators are
\begin{equation}
V\left( \left(\begin{array}{c} \psi \\ \chi \end{array} \right),z\right)=
\left(\begin{array}{cc}V(\psi,z) & W(\chi,z) \\ \overline W(\chi,z) &
V_T(\psi,z) \end{array} \right)
\end{equation}
acting on $\Hil(\Lambda)_+\oplus T$, where $\Hil(\Lambda)_+$ is the projection
of $\Hil(\Lambda)$ under the $\ze_2$ symmetry and $T$ is an irreducible module
for $\Hil(\Lambda)_+$. The operators $V(\psi,z)$ are those of $\Hil(\Lambda)$
restricted to $\Hil(\Lambda)_+$, $V_T(\psi,z)$ form a representation of
$\Hil(\Lambda)_+$ acting on $T$, while the operators $W(\chi,z)$ and
$\overline W(\chi,z)$ intertwine the two sectors.

In order that the resulting theory have a hermitian structure,
as in \reg{hermitian},
$\overline W$ is given in terms of $W$ by hermitian conjugation. Thus, we only
need to define $W$ and $V_T$ to complete the definition of the theory.

Diagrammatically, it is easier to see the situation, as in figure \ref{fig5}.
\setlength{\unitlength}{1cm}
\begin{figure}[htb]
\begin{center}
\begin{picture}(5,6)
\put(2.5,1){\oval(3,1.5)}
\put(2.5,5){\oval(3,1.5)}
\put(2,4.8){$\Hil(\Lambda)_+$}
\put(2.4,0.8){$T$}
\put(1.5,4){\vector(0,-1){2}}
\put(3.5,2){\vector(0,1){2}}
\put(0.9,2.8){$W$}
\put(3.7,2.8){$\overline W$}
\put(4.2,4.8){$V$}
\put(4.2,0.8){$V_T$}
\end{picture}
\caption{}
\label{fig5}
\end{center}
\end{figure}

$T$ is built up from a ground state (of dimension $2^{d/2}$) by the action of
creation
and annihilation operators
\begin{eqnarray}
[c^i_r,c^j_s]=r\delta_{r,-s}\delta^{ij} & r,s\in\ze+{1\over 2} & 1\leq i,j\leq
d\nonumber\\
c^i_r\chi=0\quad\forall\quad r>0 & 1\leq i\leq d & \chi\in{\rm ground\
state}\\
{c^i_r}^\dagger=c^i_{-r}\,. & &
\end{eqnarray}
The vertex operators $V_T$, forming a representation of $\Hil(\Lambda)_+$,
can be written down by analogy with the operators $V$.

Define
\begin{equation}
C^i(z)=i\sum_r{c^i_r\over r}z^{-r}\,,
\end{equation}
by analogy with $X(z)$. Then, we guess
\begin{equation}
{V_T}^G(\psi,z)=:\left(\prod^M_{a=1}{i\over(n_a-1)!}{d^{n_a}\over dz^{n_a}}
C^{i_a}(z)\right)e^{i\lambda\cdot C(z)}:\gamma_\lambda\,,
\end{equation}
where
\begin{equation}
\gamma_\lambda\gamma_\mu=(-1)^{\lambda\cdot\mu}\gamma_\mu
\gamma_\lambda\,.
\end{equation}
(The ground state forms an irreducible representation of this algebra,
introduced
since we no longer have any momentum states on which to represent the
cocycles.)

However, this is not quite correct, since in a cft it can be shown that we
require
\begin{equation}
\label{miracle}
[L_{-1},V(\psi,z)]={d\over dz}V(\psi,z)\,.
\end{equation}
We have that
\begin{equation}
L_n\equiv{1\over 2}:\sum_rc_r\cdot c_{n-r}:+{d\over 16}\delta_{n,0}
\end{equation}
satisfy the Virasoro algebra with $c=d$ (and are the modes of $V_T(\psi_L,z)$
--
which does need to be checked at the end).
So, we can check
\begin{equation}
\label{check}
[L_{-1},c^i_r]=-rc^i_{r-1}\,.
\end{equation}
Thus, $c_{1\over 2}\rightarrow c_{-{1\over 2}}$ under commutation with
$L_{-1}$, and
then normal ordering creates extra terms as the $c_{-{1\over 2}}$ is moved to
the left
past any $c_{1\over 2}$'s, for which we need to compensate.

Writing our vertex operators in an explicitly normal ordered form, {\em i.e.}
\begin{equation}
{V_T}^G(\psi,z)=\sum_{\mu\in\Lambda}\langle\mu|e^{B_-(z)}
e^{B_+(z)}|\psi\rangle\gamma_\mu\,,
\end{equation}
where
\begin{equation}
B_\pm(z)=\sum_{n\geq 0\atop r>0}B^\pm_{nr}(z)a_n\cdot c_{\pm r}\,,
\end{equation}
we see easily that
\begin{equation}
V_T(\psi,z)={V_T}^G\left(e^{A(-z)}\psi,z\right)
\end{equation}
satisfies \reg{check}, with $A(z)={1\over 2}\sum_{n,m\geq 0}A_{nm}(z)a_n\cdot
a_m$
defined uniquely up to a constant (of integration), which can be fixed by
checking the
representation property ({\em i.e.} locality of the $V_T$'s \cite{DGMtwisted}).
It is miraculous that these vertex operators turn out to be the correct ones.
We
have still to understand properly why a simple correction of the $L_{-1}$
commutation
relation should produce such far reaching consequences.

Finally, we find the intertwining operators $W$ by a trick.
The ``skew-symmetry" relation
\begin{equation}
V(\psi,z)\phi=e^{zL_{-1}}V(\phi,-z)\psi
\end{equation}
is essentially a consequence of locality and the creation axiom. In $\widetilde
\Hil(\Lambda)$, this requires
\begin{equation}
W(\chi,z)\psi=e^{zL_{-1}}V_T(\psi,-z)\chi\,,
\end{equation}
fixing $W$ (and hence $\overline W$).

To check the axioms of a cft is difficult, but has been done in
\cite{DGMtwisted},
and they are found to hold if (and only if \cite{thesis}) $\sqrt 2\Lambda^\ast$
is even, a powerful constraint as it almost forces self-duality, {\em i.e.} the
condition
for consistency of the cft on a torus, whereas we are only working on the
Riemann
sphere.
\subsection{Generalised triality}
It was observed \cite{DGMtrialsumm,DGMtwisted} from the Kac-Moody algebras
formed by the modes of the vertex operators corresponding to the states of
conformal weight one that we appeared to have $\widetilde\Hil(\Lambda_\C)
\cong\Hil(\widetilde\Lambda_\C)$ in all cases considered. This was proved
explicitly,
and we obtained the picture shown in figure \ref{fig1} (upward sloping lines
represent the untwisted constructions and downward
sloping lines the twisted
constructions).
\setlength{\unitlength}{1cm}
\begin{figure}[htb]
\begin{center}
\begin{picture}(9,5)
\put(6,4.7){$\Hil(\Lambda_\C)$}
\put(6,2.5){$\widetilde\Hil(\Lambda_\C)\cong\Hil(\widetilde\Lambda_\C)$}
\put(6,0.3){$\widetilde\Hil(\widetilde\Lambda_\C)$}
\put(3,3.6){$\Lambda_\C$}
\put(3,1.35){$\widetilde\Lambda_\C$}
\put(0.25,2.5){$\C$}
\put(0.6,2.6){\vector(2,1){2.2}}
\put(0.6,2.6){\vector(2,-1){2.2}}
\put(3.6,3.75){\vector(2,1){2.2}}
\put(3.6,3.75){\vector(2,-1){2.2}}
\put(3.6,1.5){\vector(2,1){2.2}}
\put(3.6,1.5){\vector(2,-1){2.2}}
\end{picture}
\caption{}
\label{fig1}
\end{center}
\end{figure}

The codes in fact fit together into blocks of the form
shown in figure \ref{fig2},
which are finite for 24 or more dimensions, but infinite in size for 8 and 16
dimensions.
\setlength{\unitlength}{1cm}
\begin{figure}[htb]
\begin{center}
\begin{picture}(9,5)
\put(3,3.6){$\C_1$}
\put(3,1.35){$\C_2$}
\put(3.6,3.75){\vector(2,1){2.2}}
\put(3.6,3.75){\vector(2,-1){2.2}}
\put(3.6,1.5){\vector(2,1){2.2}}
\put(3.6,1.5){\vector(2,-1){2.2}}
\multiput(4.7,4.6)(0,0.2){5}{$\cdot$}
\multiput(4.7,0.5)(0,-0.2){5}{$\cdot$}
\put(6,4.7){$\Lambda_{\C_1}$}
\put(6,2.5){$\widetilde\Lambda_{\C_1}\cong\Lambda_{\C_2}$}
\put(6,0.3){$\widetilde\Lambda_{\C_2}$}
\end{picture}
\caption{}
\label{fig2}
\end{center}
\end{figure}

Now, on $\Hil(\Lambda_\C)$ there is an obvious triality structure,
due to the existence of an affine su$(2)^d$ algebra. The isomorphism noted
above
then allows us to lift this triality up to
$\widetilde\Hil(\widetilde\Lambda_\C)$ to
give an $S_3$ group of triality automorphisms mixing straight and twisted
sectors \cite{DGMtrialsumm,DGMtwisted}.

For $\C$ the Golay code $\C_{24}$, $\widetilde\Lambda_\C$ is the Leech lattice
$\Lambda_{24}$. Now $\widetilde\Hil(\Lambda)$ inherits an obvious group of
automorphisms from Aut$(\Lambda)$ (which do not mix straight and twisted
sectors). In the case of the Leech lattice, these, together with a triality
involution,
generate the Monster group $\Monster$, which can in fact be shown to be
the full automorphism group of
$V^\natural=\widetilde\Hil(\widetilde\Lambda_{\C_{
24}})$ \cite{FLMbook,friendlygiant,Tits}.

The beauty of this picture is that
\begin{enumerate}
\item it demonstrates that the codes play a much more fundamental role than
previously thought in the cft structure, and
\item that the ``unique" features of the Monster module can be seen in a much
more general context.
\end{enumerate}
\section{Construction of Niemeier Lattices from Ternary Codes}
\label{terncodes}
Only 12 of the 24 Niemeier lattices are obtained by the two constructions from
binary
codes described in the previous section. Let us see if we can obtain more by a
study
of constructions from ternary codes \cite{PSMcodes}.

Let $\widehat\C$ be a self-dual ternary code of length $d$. The form of the
weight
enumerator, given by a theorem of Gleason \cite{ConSlo} as
\begin{eqnarray}
W_{\widehat\C}(x,y)&\in&\ce[\psi_4,\psi_{12}]\nonumber\\
\psi_4&\equiv&x(x^3+8y^3)\\
\psi_{12}&\equiv&(8x^6-160x^3y^3-64y^6)^2\,,\nonumber
\end{eqnarray}
is such that we cannot require it to be the equivalent of doubly-even, {\em
i.e.}
${\hat c}^2\in 3\ze$ $\forall$ $\hat c\in\widehat\C$ but we cannot require
${\hat c}^2\in 6\ze$ $\forall$ $\hat c\in\widehat\C$.

So, we define
\begin{eqnarray}
\widehat\C_+&=&\left\{\hat c\in\widehat\C:{\rm wt}(\hat c)\equiv{0\bmod
6}\right\}\nonumber\\
\widehat\C_-&=&\left\{\hat c\in\widehat\C:{\rm wt}(\hat c)\equiv{3\bmod
6}\right\}\,.\\
\end{eqnarray}
Set
\begin{equation}
\Lambda_0(\widehat\C)=\left({\widehat\C_+\over\sqrt 3}+\sqrt
3\ze^d_+\right)\cup
\left({\widehat\C_-\over\sqrt 3}+\sqrt 3\ze^d_-\right)\,.
\end{equation}
This is even, but not self-dual. We add to it a sector shifted by some vector,
{\em i.e.}
set
\begin{equation}
\Lambda_\pm(\widehat\C)=\left({\widehat\C_++{3\over 2}\underline 1\over\sqrt
3}+\sqrt 3\ze^d_\pm\right)\cup
\left({\widehat\C_-+{3\over 2}\underline 1\over\sqrt 3}+\sqrt
3\ze^d_\mp\right)\,.
\end{equation}
Then
$\Lambda_{\widehat\C}^\pm=\Lambda_0(\widehat\C)\cup\Lambda_\pm(\widehat\C)$
is an even self-dual lattice, provided $\underline 1\in\widehat\C$ and
$d\in24\ze$.
Note that $\Lambda_{\widehat\C}^\pm$ can be regarded as the two possible
results
of applying just one of the constructions to the set of codes equivalent to
$\widehat\C$.

Note that we already have all the even self-dual lattices from binary codes in
8
and 16 dimensions, so the restriction to $d\in24\ze$ is not too disastrous. Let
us
now consider the case $d=24$ exclusively.

{}From the results of Venkov \cite{Venkov}, we may identify the resulting
Niemeier
lattice uniquely by the number of vectors of squared length two and the number
of orthogonal components into which the set of such vectors splits.

In fact, we find
\begin{equation}
\left|\Lambda_{\widehat\C}^\pm(2)\right|=3n_3+n_6+n_{24}^\pm\,,
\end{equation}
where $n_m$ is the number of codewords of weight $m$, and $n_m^\pm$ is the
number of codewords of weight $m$ in which the number of entries equal to $1$
is (an) even or odd (multiple of 3) respectively.

However, for the purposes of computer calculation, this naive approach results
in
considerably too much computation. So, we use a theorem of Mallows, Pless and
Sloane \cite{MallPlessSloan} on the form of
the complete weight enumerator of a self-dual
ternary code (similar to Gleason's result for the weight enumerator),
and we find that there are $\left|\Lambda_{\widehat\C}^+(2)\right|/24-2$
codewords of weight 6 with 6 1's in the code, with a similar result for
$\left|\Lambda_{\widehat\C}^-(2)\right|$.

We now need some codes to work on. However, the self-dual ternary codes have
only been classified up to and including length 20. (There are 24 of these!) We
do not need a full classification however, since it is really only the
resulting lattices
in which we are interested.

We proceed in two steps. Firstly, we show that we may construct a self-dual
length
24 code with $n_3\geq 4$ from a self-dual length 20 code (see \cite{PSMcodes}
for full details).
The construction involves
a choice of codeword of weight 18 in the length 20 code. The support of this
codeword
is all that is relevant, since sign changes in the coordinates trivially give
rise to
equivalent codes of length 24. As a bonus, we obtain an upper bound on the
number
of inequivalent length 24 self-dual ternary codes, though quite a bit larger
(of order
1000) than the lower bound of 40 found by Leon, Pless and Sloane \cite{LPS}.

One can show, for $n_3\geq 2$, that $\Lambda_{\widehat\C}^+\cong
\Lambda_{\widehat\C}^-$, and we find, by our previously described calculational
technique, the corresponding lattices, 19 of them in total (11 of the 12 not
produced
by the binary constructions -- the exception being $A_{24}$). This suggests
that
we should try to find {\em all} of the Niemeier lattices by these
constructions, not
merely those which cannot be obtained by the binary constructions.

[As an aside, we find (distinguishing them by the lattice they produce as well
as
their complete weight enumerators) at least 34 inequivalent indecomposable
codes,
{\em c.f.} the lower bound of 13 given by  Leon, Pless and Sloane.]

There are some known length 24 codes containing $\underline 1$ with $n_3<4$.
Using
these, we get 23 of the 24 Niemeier lattices. In particular, we get the Leech
lattice from
at least two codes ($Q_{24}$ and $P_24$), which is surprising as we would
normally
expect some sort of uniqueness to be associated to this structure.

Finally, we resort to a random basis generation technique on a computer, and
find
the $A_{24}$ Niemeier lattice.

So, we obtain all Niemeier lattices from the constructions
$\Lambda_{\widehat\C}^\pm$,
23 of them from $\Lambda_{\widehat\C}^+$ alone
($\left|\Lambda_{\widehat\C}^+(2)\right|
\geq 48$ by the known form of the
complete weight enumerator, and so the Leech lattice cannot be
obtained) and at least 22 from $\Lambda_{\widehat\C}^-$.

As yet, there is no analogous cft construction. It would seem to have to be a
$\ze_2$-orbifold of a $\ze_2$-orbifold, and we only know explicitly how to
orbifold
the theories $\Hil(\Lambda)$ so far, though one can make arguments about
orbifolds
of other theories without knowledge of their construction by looking at the
modular
transformations of the generalised Thompson series \cite{PSMSchell}. This is a
way
of tackling the classification problem for $c=24$ theories which will be
discussed
further in section \ref{class}.
\section{$Z\!\!\!Z_3$-Orbifolds and Ternary Codes}
\label{ze3}
In the absence of any inspiration for an analogue of section \ref{terncodes}
for cft's,
let us bring in ternary codes into the picture in another way. The following in
fact applies
to any automorphism of prime order $p$ with $(p-1)|24$, {\em i.e.} $p=3$, 5, 7
and 13.
We shall restrict to $p=3$ simply for ease of exposition, and consider a
$\ze_3$-NFPA
(no-fixed-point automorphism) orbifold of $\Hil(\Lambda)$.
\subsection{Reformulation of $\Hil(\Lambda)$}
We begin with the theory $\Hil(\Lambda)$ for $\Lambda$ an even self-dual
lattice of
dimension $2d$. As remarked before, any automorphism $R$ of $\Lambda$ extends
to an  automorphism of the cft, {\em i.e.}
\begin{eqnarray}
a_n^i&\mapsto&{R^i}_ja^j_n\nonumber\\
|\lambda\rangle&\mapsto&(-1)^{\lambda\cdot\mu}|R\lambda\rangle\,,\qquad
\mu\in\Lambda/2\Lambda\,.
\end{eqnarray}
[In fact, it has been shown that any finite order automorphism of
$\Hil(\Lambda)$
is conjugate to such an automorphism \cite{PSMorb}.]

Now, suppose that $\Lambda$ admits a third order NFPA. It is then easily shown
that $\Lambda$ corresponds to a $d$-dimensional complex lattice
$\widehat\Lambda$
over the Eisenstein ring $\Epsilon$ of cyclotomic integers $\ze[\omega]=\{
m+n\omega:m,n\in\ze\}$, $\omega=e^{2\pi i/3}$. Let us rewrite the theory
$\Hil(\Lambda)$
in terms of oscillators $b_n^i$, $\overline b_n^i$, $1\leq i\leq d$, such that
\begin{eqnarray}
\ [b_m^i,\overline b_n^j]&=&m\delta_{m,-n}\delta^{ij}\nonumber\\
\ [b_m^i,b_n^j]=&0&=[\overline b_m^i,\overline b_n^j]\nonumber\\
{b^i_n}^\dagger&=&\overline b_{-n}^i\\
(p\equiv b^i_0)|\lambda\rangle&=&{\lambda\over\alpha}|\lambda\rangle\,,\qquad
\lambda\in\widehat\Lambda\nonumber\\
(\overline p\equiv\overline b^i_0)|\lambda\rangle&=&
{\overline\lambda\over\alpha}|\lambda\rangle\nonumber\\
b^i_n|\lambda\rangle=\overline b^i_n|\lambda\rangle&=&0\,,\qquad
n>0\,,\nonumber
\end{eqnarray}
for some scale factor $\alpha$, fixed later by locality.

For
\begin{equation}
\psi=\prod_{a=1}^Mb^{i_a}_{-m_a}\prod_{b=1}^N\overline b^{j_b}_{-n_b}
|\lambda\rangle\,,
\end{equation}
we have
\begin{equation}
V(\psi,z)=:\prod_{a=1}^M{i\over(m_a-1)!}{d^{m_a}\over dz^{m_a}}X^{i_a}_+(z)
\prod_{b=1}^N{i\over(n_b-1)!}{d^{n_b}\over dz^{n_b}}X^{j_b}_-(z)
e^{i{\overline\lambda\over\alpha}\cdot X_+(z)}e^{i{\lambda\over\alpha}\cdot
X_-(z)}:
\sigma_\lambda\,,
\end{equation}
where
\begin{equation}
X_+(z)=q-ip\ln z+i\sum_{n\neq 0}{b_n\over n}z^{-n}
\end{equation}
and
\begin{equation}
X_-(z)=\overline q-i\overline p\ln z+i\sum_{n\neq 0}{\overline b_n\over
n}z^{-n}\,,
\end{equation}
with
\begin{equation}
e^{i{\overline\lambda\over\alpha}\cdot q+(z)}e^{i{\lambda\over\alpha}\cdot
\overline q}|\mu\rangle=|\lambda+\mu\rangle\,.
\end{equation}
The relation
\begin{equation}
\hat\sigma_\lambda(\equiv e^{i{\overline\lambda\over\alpha}\cdot
q+(z)}e^{i{\lambda\over\alpha}\cdot
\overline q})\hat\sigma_\mu=(-1)^{\langle\lambda,\mu\rangle/\alpha^2}
\hat\sigma_\mu\hat\sigma_\lambda
\end{equation}
ensures locality.

The third order automorphism $\theta$ in this picture is simply given by
\begin{eqnarray}
\theta b_n^i\theta^{-1}&=&\omega b_n^i\nonumber\\
\theta\overline b^i_n\theta^{-1}&=&\overline\omega\overline b^i_n\\
\theta|\lambda\rangle&=&|\overline\omega\lambda\rangle\nonumber
\end{eqnarray}
\subsection{Construction of $\widehat\Hil(\Lambda)$}
We now need to define the other vertex operators in the orbifold theory. We
shall
have three sectors (one for each conjugacy class of the discrete symmetry group
\cite{Mythesis}). In diagrammatic form we have figure \ref{fig6}.
\setlength{\unitlength}{1cm}
\begin{figure}[htb]
\begin{center}
\begin{picture}(5,6)
\put(3,1){\oval(2,1)}
\put(3,3){\oval(2,1)}
\put(3,5){\oval(2,1)}
\put(2.9,4.8){0}
\put(2.9,2.8){1}
\put(2.9,0.8){2}
\put(2.5,4.4){\vector(0,-1){0.8}}
\put(3.5,3.6){\vector(0,1){0.8}}
\put(2.5,2.4){\vector(0,-1){0.8}}
\put(3.5,1.6){\vector(0,1){0.8}}
\put(1.8,3.8){$W_1$}
\put(1.8,1.8){$W_3$}
\put(3.7,3.8){$\overline W_1$}
\put(3.7,1.8){$\overline W_3$}
\put(4.2,0.8){$V_2$}
\put(4.2,2.8){$V_1$}
\put(4.2,4.8){$V_0$}
\put(1,5){\vector(0,-1){4}}
\put(5,1){\vector(0,1){4}}
\put(0.3,2.8){$W_2$}
\put(5.2,2.8){$\overline W_2$}
\end{picture}
\caption{}
\label{fig6}
\end{center}
\end{figure}

Sector 0 is the $\theta=1$ projection of $\Hil(\Lambda)$, while sectors 1 and 2
form irreducible meromorphic representations of sector 0. So, we write down
vertex operators $V_1(\psi,z)$ acting on sector 1, $V_2(\psi,z)$ acting on
sector 2,
for $\psi$ a state in sector 0, by analogy with $V_0(\psi,z)$, just as in the
$\ze_2$
case.

Introduce oscillators $c_r^i$, $1\leq i\leq d$, $r\in\ze\pm{1\over 3}$, such
that
\begin{equation}
[c^i_r,c^j_s]=r\delta_{r,-s}\delta^{ij}\,,\qquad {c^i_r}^\dagger=c^i_{-r}\,,
\end{equation}
and similarly operators $\overline c^i_r$. All other commutation relations
vanish.
Sector 1 is then built up by the action of the $c$'s on some ground state, and
sector
2 similarly by the $\overline c$'s.

The operators
\begin{equation}
L_n=\sum_{r\in\ze+{1\over 3}}:c_r\cdot c_{n-r}:+{d\over 9}\delta_{n0}
\end{equation}
satisfy the $c=2d$ Virasoro algebra (and turn out, as necessary for
consistency, to
be the modes of $V_1(\psi_L,z)$). The conformal weight of
\begin{equation}
\chi=\prod_{a=1}^Mc^{j_a}_{-r_a}\chi_0\,,
\end{equation}
for $\chi_0$ in the ground state, is then ${d\over 9}+\sum_{a=1}^Mr_a$. Hence,
for a
meromorphic representation, we require $d\in3\ze$ and we must project onto
states
with $\theta=1$. where
\begin{eqnarray}
\theta c^i_r\theta^{-1}&=&e^{2\pi ir}c^i_r\nonumber\\
\theta\chi_0&=&e^{-2\pi id/9}\chi_0
\end{eqnarray}
extends the automorphism from sector 0 to sector 1 (provided we define $V_1$
suitably).

Set
\begin{equation}
V_1^G(\psi,z)=:\prod_{a=1}^M{i\over(m_a-1)!}{d^{m_a}\over dz^{m_a}}C^{i_a}_+(z)
\prod_{b=1}^N{i\over(n_b-1)!}{d^{n_b}\over dz^{n_b}}C^{j_b}_-(z)
e^{i{\overline\lambda\over\alpha}\cdot C_+(z)}e^{i{\lambda\over\alpha}\cdot
C_-(z)}:
\gamma_\lambda\,,
\end{equation}
where
\begin{equation}
C_\pm(z)=i\sum_{r\in\ze\pm{1\over 3}}{c_r\over r}z^{-r}
\end{equation}
and the $\gamma_\lambda$'s are a set of cocycle operators (for which the ground
state forms an irreducible representation, though we will not go into such
details
here). This definition is simply an analogy of that for $V_0$ in terms of
$X_\pm$,
{\em c.f.} the $\ze_2$ case. (Note that $\theta V^G_1((\psi,z)\theta^{-1}
=V_1^G(\theta\psi,z)$.)

Similarly, we define $V_2^G$ by replacing $C_\pm$ by $\overline C_\mp$, where
these
have the obvious definition.

However, as in the $\ze_2$ case, there is a problem, in that the relation
\reg{miracle} is not satisfied by $V_1^G$, $V_2^G$ due to a similar normal
ordering problem to that in the $\ze_2$ case. We find, as before, that this can
be corrected by setting
\begin{equation}
V_1(\psi,z)=V_1^G\left(e^{A_1(-z)}\psi,z\right)\,,
\end{equation}
where
\begin{equation}
A_1(z)=\sum_{n,m\geq 0}A^1_{nm}(z)b_n\cdot\overline b_m
\end{equation}
and the $\ce$-numbers $A^1_{nm}$ are determined up to one arbitrary
constant (of integration) fixed later by locality requirements.
A similar result holds for $V_2(\psi,z)$, {\em i.e.} the expression written
down
by analogy with sector 0 is ``almost" correct. Again, we still have no proper
understanding of why this procedure works.

So, we have defined the representations of sector 0. Now we need to define
the intertwining operators mixing the three sectors. Also, of course, if
we want to show we have a cft we must verify all the locality
relations. For the sake of brevity , however, let us simply content
ourselves with defining the operators
$W_1$, $W_2$, $W_3$ and $\overline W_1$, $\overline W_2$, $\overline W_3$.
Full details of the locality calculations may be found in \cite{PSMthird}.

In the original theory $\Hil(\Lambda)$ we have \reg{hermitian},
and if we are to preserve such a relation in the orbifold theory we must have
\begin{equation}
\overline W_i(\overline\chi,z)^\dagger=W_i\left(e^{z^\ast L_1}{z^\ast}^{-2L_0}
\chi,1/z^\ast\right)\,,
\end{equation}
extending the conjugation map to interchange the two twisted sectors.
This is just as in the $\ze_2$ case.

$W_1$ and $W_2$ are defined easily by the same trick as before, {\em i.e.} we
must
have the locality relation
\begin{equation}
W_i(\chi,z)V_0(\psi,w)=V_i(\psi,w)W_i(\chi,z)\,,
\end{equation}
for $i=1$, 2. Acting on the vacuum state and requiring the creation property
$V(\rho,z)|0\rangle=e^{zL_{-1}}|\rho\rangle$ for the orbifold cft gives us
 \begin{equation}
W_i(\chi,z)e^{wL_{-1}}|\psi\rangle=V_i(\psi,w)e^{zL_{-1}}|\chi\rangle\,,
\end{equation}
or
\begin{equation}
W_i(\chi,z-w)|\psi\rangle=e^{(z-w)L_{-1}}V_i(\psi,w-z)|\chi\rangle\,.
\end{equation}
Hence
\begin{equation}
W_i(\chi,z)|\psi\rangle=e^{zL_{-1}}V_i(\psi,-z)|\chi\rangle\,,
\end{equation}
for $i=1$, 2, fixes $W_i$ uniquely. (Note that we here effectively derived the
``skew-symmetry" relation which we quoted at the corresponding point
in section \ref{ze2}, giving a flavour of the sort of manipulations involved.)
Though it remains to be checked that this definition is consistent with
locality, it is certainly forced upon us as a consequence of locality.
\subsubsection{Definition of $W_3$}
Defining $W_3$ is more complicated. This is the point at which we must leave
the analogy with the $\ze_2$ case.
It maps from sector 1 to sector 2, and
so we cannot use the above trick of allowing it to act on the vacuum.
Instead, we use the fact that sectors 1 and 2 are irreducible representations
of sector 0 to give an implicit definition.

Fix $\chi_0\in$ sector 1. Then define
\begin{equation}
W_3(V_1(\phi_1,w_1)\chi_0,z)V_1(\phi_2,w_2)\chi_0=
V_2(\phi_2,w_2)V_2(\phi_1,w_1+z)W_3(\chi_0,z)\chi_0\,,
\end{equation}
which must be true if we are to have locality in the orbifold theory, {\em
i.e.}
$W_3$ is defined in terms of $W_3(\chi_0,z)\chi_0\equiv F_{\chi_0}(z)$ for some
fixed state $\chi_0$.
[We lose the advantage over previous work of having an explicit form
for our vertex operators by such
an approach, while also obscuring the symmetry between $W_3$ and
$\overline W_3$ under $c\leftrightarrow\overline c$, though this has been
demonstrated \cite{thesis}.]

Take $\chi_0$ as a ground state for simplicity (and work with $c$ a multiple of
72).
Then let us conjecture
\begin{equation}
F_{\chi_0}(z)=z^{-\Delta_0}\sum_\alpha e^{\sum_{s>0}G_sz^s\alpha\cdot\overline
c_{-s}}e^{\sum_{r>0}F_rz^r\overline \alpha\cdot\overline
c_{-r}}e^{\sum_{r,s>0}D_{rs}z^{r+s}\overline c_{-r}\cdot\overline c_{-s}}
\chi_\alpha(\chi_0)\,,
\end{equation}
where $\Delta_0$ is the conformal weight of the twisted sector ground state,
$r\in\ze+{1\over 3}$, $s\in\ze-{1\over 3}$, $\chi_\alpha$ is a ground state and
the sum over $\alpha$ is over some lattice (presumably either $\Lambda$
or its dual).

Now, since $\chi_0$ is {\em quasi-primary} (annihilated by $L_1$),
\begin{equation}
e^{{1\over z}L_1}F_{\chi_0}(\zeta)=\left(1-{\zeta\over z}\right)^{-2\Delta_0}
F_{\chi_0}\left({z\zeta\over z-\zeta}\right)\,,
\end{equation}
by the standard M\"obius transformation result (or hermitian conjugation
of the $L_{-1}$ commutation relation, if preferred).
This gives us
\begin{equation}
F_r={F\over r}\left({-{2\over 3}\atop r-{1\over 3}}\right)(-1)^{r-{1\over
3}}\,,
\end{equation}
and similarly for $G_s$ up to some constant $G$, while
\begin{equation}
(r+s)D_{r,s}=(r+1)D_{r+1,s}+(s+1)D_{r,s+1}+{2\over 9}D_{{1\over 3},s}D_{
r,{2\over 3}}\,.
\end{equation}
The constants $F$ and $G$ are easily determined by trying to check a suitable
locality
relation ($G=0$ in fact). Solving for the coefficients $D_{rs}$ is however more
of a problem. We have one indeterminate at each level (level$(D_{rs})\equiv
r+s$).

Now, in the case of $W_1$ and $W_2$, the skew-symmetry relation fixed
the operators uniquely. In this case, it gives
\begin{equation}
e^{zL_{-1}}F_{\chi_0}(-z)=F_{\chi_0}(z)\,.
\label{cath}
\end{equation}
Let us also consider the locality relation
\begin{equation}
\overline W_2(\chi_0,z)F_{\chi_0}(\zeta)=\overline
W_2(\chi_0,\zeta)F_{\chi_0}(z)\,.
\end{equation}
Replace $A_{-m}$ by $x^m$ and $\overline a _{-n}$ by $y^n$ in these equations
to obtain a functional relation. (This is actually also sufficient in proving
locality,
since we know that we only have exponentials of bilinears, whereas in general
it would be impossible in  reversing the argument to identify {\em e.g.} $x^2$
with $a_{-1}\cdot a_{-1}$ or $a_{-2}$ uniquely.)
Set
\begin{equation}
g(a,b)\equiv\sum_{r,s>0}D_{rs}a^rb^s\,.
\end{equation}
Then we find that these two relations, after much manipulation, reduce to
\begin{equation}
g(a,b)-g(1-a,1-b)=-(1-\omega)\ln\left({(1-a)^{1\over 3}-\omega(1-b)^{1\over
3}\over
(-a)^{1\over 3}-\omega(-b)^{1\over
3}}\right)+(\omega\rightarrow\overline\omega)
\end{equation}
and
\begin{equation}
g(a,b)-g\left({a\over a-1},{b\over b-1}\right)=-(1-\omega)\ln\left({(1-{1\over
b})^{1\over 3}
-\omega(1-{1\over a})^{1\over 3}\over
({1\over b})^{1\over 3}-\omega({1\over a})^{1\over 3}}\right)
+(\omega\rightarrow\overline\omega)\,.
\end{equation}
Now, we only need one relation at each level, so consider these for $a=b$.
Also, consider
\begin{equation}
h(a)\equiv\left.{d\over da}{d\over db}g(a,b)\right|_{a=b}
\end{equation}
to remove the logarithms ($h$ arises in correlation functions, and so is a more
natural
object to consider in any case).

We find
\begin{equation}
h(a)-h(1-a)={2a-1\over 9a^2(1-a)^2}
\end{equation}
and
\begin{equation}
h(a)-{1\over(1-a)^4}h\left({a\over a-1}\right)={2-a\over 9a(1-a)^2}\,.
\end{equation}
Noting an obvious solution to these, write
\begin{equation}
h(a)={1\over 9a(1-a)^2}+k(a)\,,
\end{equation}
where
\begin{equation}
k(a)=k(1-a)={1\over(1-a)^4}k\left({a\over a-1}\right)\,.
\label{link}
\end{equation}
Now, $h(a)$ has a simple pole at the origin, from its definition, of residue
${2\over 9}D_{{1\over 3}{2\over 3}}$. But the skew-symmetry relation \reg{cath}
gives us $D_{{1\over 3}{2\over 3}}={1\over 2}$, and hence $k(a)$ is regular at
0, and
therefore at 1 and $\infty$ by \reg{link}. Since $h$ is essentially a
correlation function,
these are the only possible poles. So Liouville's theorem tells us that $k$ is
constant, and furthermore this constant value vanishes by \reg{link}.

We can then evaluate $D_{rs}$ at successive levels. Doing so for the first
half dozen or so, one arrives at the (surprisingly asymmetric) conjecture
\begin{equation}
D_{rs}={(-1)^{r+s}\over r+s}\left({1\over 3s}-1\right)\left({-{2\over 3}\atop
r-{1\over 3}}
\right)\left({-{1\over 3}\atop s-{2\over 3}}
\right)\,,
\end{equation}
which is found, by substitution, to be the solution.

Note however that full verification of the consistency of the
$\ze_3$-orbifold cft has yet to be completed.
\subsection{Construction of Eisenstein lattices from ternary codes}
Now, we may generalise a pair of constructions due to Sloane of
$\Epsilon$-lattices from ternary codes, inspired by results such as
those shown in figure \ref{fig3},
which strongly indicate, having seen the binary situation, the
existence of some code structure to the left, presumably based on a
ternary code [and also that the cft's are consistent!].
\setlength{\unitlength}{1cm}
\begin{figure}[htb]
\begin{center}
\begin{picture}(8,9)
\put(5.1,4.7){$\Hil({A_2}^{12})$}
\put(5.1,2.5){$\Hil(\Lambda_{24})$}
\put(5.1,0.3){$V^\natural\ ?$}
\put(2,3.7){${A_2}^{12}$}
\put(2,1.35){$\Lambda_{24}$}
\put(2,5.9){${E_6}^4$}
\put(2.7,3.75){\vector(2,1){2.2}}
\put(2.7,3.75){\vector(2,-1){2.2}}
\put(2.7,1.5){\vector(2,1){2.2}}
\put(2.7,1.5){\vector(2,-1){2.2}}
\put(2.7,6.05){\vector(2,1){2.2}}
\put(2.7,6.05){\vector(2,-1){2.2}}
\put(5.1,7.05){$\Hil({E_6}^4)$}
\put(2,8.2){Lattice}
\put(5.1,8.2){CFT}
\end{picture}
\caption{}
\label{fig3}
\end{center}
\end{figure}

Let $\widehat\C$ be a self-dual ternary code of length $d$. Define the
``straight'' construction by
\begin{equation}
\Lambda_\Epsilon(\widehat\C)=\widehat\C+(\omega-\overline\omega)\Epsilon^d\,,
\end{equation}
and for $\underline 1\in\widehat\C$ define the ``twisted''
construction by
\begin{equation}
\widetilde\Lambda_\Epsilon(\widehat\C)=\Lambda_0(\widehat\C)\cup
\Lambda_1(\widehat\C)\cup\Lambda_2(\widehat\C)\,,
\end{equation}
where
\begin{eqnarray}
\lambda_0(\widehat\C)&=&\widehat\C+(\omega-\overline\omega)\Epsilon^d_0\nonumber\\
\lambda_1(\widehat\C)&=&\widehat\C+(\omega-\overline\omega)
\left(\Epsilon^d_D+{1\over 3}\underline 1\right)\\
\lambda_2(\widehat\C)&=&\widehat\C+(\omega-\overline\omega)
\left(\Epsilon^d_{-D}-{1\over 3}\underline 1\right)\nonumber\,,
\end{eqnarray}
with
\begin{equation}
\Epsilon^d_\rho\equiv\left\{x=(x_1,\ldots,x_d)\in\Epsilon^d:
\sum_{i=1}^dx_i\equiv{\rho\bmod(\omega-\overline\omega)}\right\}
\end{equation}
and $d=12D$.
\subsection{Results and relationships between the constructions}
For $D=1$, we have 3 self-dual codes of length 12. Two contain
$\underline 1$, and so can be used for the twisted construction.

We identify the corresponding Niemeier lattices as before, and we
obtain figure \ref{fig4},
\setlength{\unitlength}{1cm}
\begin{figure}[htb]
\begin{center}
\begin{picture}(9,12)
\put(6.8,7){$\Hil({E_6}^4)$}
\put(6.8,0.3){$V^\natural$\ ?}
\put(6.8,2.5){$\Hil(\Lambda_{24})$}
\put(6.8,4.75){$\Hil({A_2}^{12})$}
\put(3.7,3.7){${A_2}^{12}$}
\put(3.7,1.35){$\Lambda_{24}$}
\put(3.7,5.85){${E_6}^4$}
\put(0.4,2.5){$\C_{12}$}
\put(-0.15,4.75){$4\C_3(12)$}
\put(1.3,2.6){\vector(2,1){2.2}}
\put(1.3,2.6){\vector(2,-1){2.2}}
\put(1.3,4.9){\vector(2,1){2.2}}
\put(1.3,4.9){\vector(2,-1){2.2}}
\put(4.4,3.75){\vector(2,1){2.2}}
\put(4.4,3.75){\vector(2,-1){2.2}}
\put(4.4,1.5){\vector(2,1){2.2}}
\put(4.4,1.5){\vector(2,-1){2.2}}
\put(4.4,6){\vector(2,1){2.2}}
\put(4.4,6){\vector(2,-1){2.2}}
\put(6.8,8.4){$\Hil({A_8}^3)$}
\put(6.8,10.7){$\Hil({E_8}^3)$}
\put(3.7,9.6){${E_8}^3$}
\put(0.3,9.6){$3\C_4$}
\put(1.3,9.7){\vector(1,0){2.2}}
\put(4.4,9.7){\vector(2,1){2.2}}
\put(4.4,9.7){\vector(2,-1){2.2}}
\end{picture}
\caption{}
\label{fig4}
\end{center}
\end{figure}
confirming that we do indeed have the correct analogy. (We identify the twisted
cft's by use of a theorem in \cite{DGMtriality}, which states that when the
rank of
the Lie algebra generated by the zero modes of the vertex operators
corresponding to
the states of conformal weight one is equal to the central charge, the cft is
isomorphic
to a cft of the form $\Hil(\Lambda)$, for some even lattice $\Lambda$.)

In some sense, the constructions of $\ze$-lattices in the previous
section
are $d$-dimensional projections of the constructions of
$2d$-dimensional
$\ze$-lattices from ternary codes described here. So, we would guess
that all $c=24$ self-dual cft's could be obtained by projecting out
the $c=48$ cft's produced by the straight and $\ze_3$-twisted
constructions on suitable 24-dimensional Eisenstein lattices. The
status of this remark is as yet unclear.

In any case, we must verify that
$\widehat\Hil(\Lambda_\Epsilon(\C_{12}))$
is $V^\natural$. We provide evidence for this only. One point is that
the ternary code $\C_{12}$ enjoys similar unique properties to the
binary Golay code $\C_{24}$. Also, $\Lambda_\Epsilon(\C_{12})$, the
complex Leech lattice, has symmetry group 6Suz. Note that ${\rm
F}_{3-}=$Suz, {\em c.f.} ${\rm F}_{2-}={\rm Aut}\
(\Lambda_{24})/\ze_2$, suggesting that then automorphism group is
again $\Monster$. (We will discuss the uniqueness of the Monster
module in the next section.)
An analogous analysis to that for the $\ze_2$ case should complete the
Suzuki group to the Monster and also generalise to other codes. [We
utilise the affine su$(3)^d$ algebra present in $\Hil(\Lambda_\Epsilon
(\widehat\C))$.]
\section{Classification of Self-Dual $c=24$ Conformal Field Theories
and Uniqueness of $V^\natural$}
\label{class}
Finally, we will briefly summarize and discuss some results on the
classification
of self-dual cft's with $c=24$. Schellekens \cite{Schell:Venkov}
has derived results for these analogous to those of Venkov \cite{Venkov}
for the Niemeier lattices.
In particular, he has shown that the Kac-Moody algebra generated by the modes
of the
states of conformal weight one is restricted to contain components whose
central
charges (under the Sugawara construction -- see {\em e.g.} \cite{GoddOl})
sum to 24 and have
\begin{equation}
{g\over k}={N\over 24}-1\,,
\end{equation}
where $N$ is the number of weight one states in the cft, $g$ is the Coxeter
number
of the Kac-Moody component and $k$ is its level.
[He has since derived stronger constraints \cite{SchellComplete}, reducing to
71
(the largest prime divisor of the order of the Monster!)
the number of possible Kac-Moody algebras, though only 39 have been found
in constructed cft's so far (those constructed in section \ref{ze2}). Also,
there are no known examples as yet of distinct cft's with the same algebra,
though
some do have coincident partition functions.]

Let us look at the consequences of this work for the uniqueness of a self-dual
$c=24$ theory $\Hil$ with no weight one states.

Suppose there exists an involution $g$ (say if $\Hil$ were a Monster module,
or an orbifold of $V^\natural$ with respect to some known automorphism as
considered in recent work by Tuite \cite{MichaelHep}), and consider the
orbifold
$\Hil_g$ constructed using it (which we shall assume to exist). We then have
the
partition function (using notation explained in {\em e.g.} \cite{Ginsparg})
\begin{eqnarray}
\chi_{\Hil_g}(\tau)&=&{1\over 2}\left(
{1
\lower 6pt \hbox{$
{\square\atop{\displaystyle 1}}$}}(\tau)+
{g
\lower 6pt \hbox{$
{\square\atop{\displaystyle 1}}$}}(\tau)+
{1
\lower 6pt \hbox{$
{\square\atop{\displaystyle g}}$}}(\tau)+
{g
\lower 6pt \hbox{$
{\square\atop{\displaystyle g}}$}}(\tau)\right)\nonumber\\
&\equiv&{1\over 2}\left(J(\tau)+T_g(\tau)+T_g(S(\tau))+T_g(ST(\tau))\right)\,.
\end{eqnarray}
Now, we proceed along the lines of work done by Tuite \cite{Tuite:moon}.
Clearly $T_g(\tau)$ is $\Gamma_0(2)$ invariant. If $\Hil_g$ has the correct
twisted sector ground state energy ($\geq 1$) then $T_g(\tau)$ is a
$\Gamma_0(2)$
hauptmodul, and so is known, {\em i.e.}
\begin{equation}
T_g(\tau)=\left({\eta(\tau)\over\eta(2\tau)}\right)^{24}+24\,.
\end{equation}
This then gives
\begin{equation}
\chi_{\Hil_g}(\tau)=J(\tau)+24\,.
\end{equation}
Schellekens' partial classification then restricts us to an algebra
U$(1)^{24}$. Now, as mentioned previously, we have the theorem that
when the rank of the corresponding Lie algebra is equal to the central
charge, a cft $\Hil\cong\Hil(\Lambda)$ for some even lattice $\Lambda$.
We thus deduce $\Hil_g\cong\Hil(\Lambda_{24})$ (the Leech lattice being the
only lattice producing the appropriate partition function).

To proceed analogously to Venkov's proof of the uniqueness of the Leech
lattice, we would have to assume the existence of an orbifolding inverse to the
original,
{\em i.e.} such that $(\Hil_g)_h\equiv\Hil$.

Instead, we write $\Hil_g={\Hil_g}^0\oplus U$, where ${\Hil_g}^0$ is the
subspace
of $\Hil$ invariant under the action of $g$ and $U$ forms an irreducible
representation
of ${\Hil_g}^0$.

Now, $a_{-1}^i|0\rangle\in U$, so $a_{-m}^ia^j_{-n}|0\rangle$ lie in
${\Hil_g}^0$, as
$U\times U\subset{\Hil_g}^0$. So ${\Hil_g}^0\cong\Hil(\Lambda_{24})_+$
(consistent
with its known partition function). But there exist only 2 irreducible
representations
of $\Hil(\Lambda)_+$ for $\Lambda$ self-dual. These are easily distinguished by
the
number of weight one states, and we deduce that $\Hil\cong V^\natural$, as
required.

Note that we may also, of course, use a similar argument beginning with an
automorphism of higher order, though some of the relevant results remain to be
verified in those cases.
\section{Conclusions}
To conclude, we have shown how the connections between $\ze_2$-orbifolds
and binary codes lead to a generalization of the triality structure of FLM,
and that similar links between ternary codes and $\ze_3$-orbifolds appear
to generalise another ``triality" to a class of conformal field theories.
The orbifold construction should follow through as in the $\ze_3$ case
for higher prime ordered twists, providing a series of constructions of the
Monster module. In addition, the construction of the Niemeier lattices from
ternary
codes holds out the hope that a conformal field theory analogue will produce a
complete set of self-dual theories, at least at central charge 24. This is work
which
is currently in progress. The comments regarding the uniqueness of the Monster
module (or rather a self-dual theory with no weight one states) are intimately
connected with the ideas of Tuite. Much research is still to be done in this
promising area on the boundary between fundamental physics and the hitherto
abstract realms of groups and the theory of modular functions.

\end{document}